# Resistive switching characteristics of Cu/MgO/MoS$_2$/Cu structure[*]

HE Xiaolong, CHEN Peng

School of Physics Science and Technology, Southwest University, Chongqing 400715, China

**Abstract** During the study of resistive switching devices, researchers have found that the influence of the insertion layer cannot be ignored. Many reports have confirmed that the appropriate insertion layer can significantly improve the performance of the resistive switching devices. Therefore, in this work, we use magnetron sputtering to fabricate three devices: Cu/MgO/Cu, Cu/MgO/MoS$_2$/Cu and Cu/MoS$_2$/MgO/Cu. Through the characterization test of each device and the measurement of the *I-V* curve, it is found that the resistive switching characteristics of the Cu/MgO/Cu device will change greatly after adding an MoS$_2$ insertion layer. The analysis results show that the inserted MoS$_2$ layer does not change the main transmission mechanism (space charge limited conduction) of the device, but affects the regulating function of interfacial potential barrier, the effect also is related to the location of MoS$_2$ inserted into the layer. Among the Cu/MgO/Cu, Cu/MgO/MoS$_2$/Cu and Cu/MoS$_2$/MgO/Cu devices, the Cu/MgO/MoS$_2$/Cu device exhibits a larger switching ratio (about 10$^3$) and a lower reset voltage (about 0.21 V), which can be attributed to the regulation of the interface barrier between MgO and MoS$_2$. In addition, when the MoS$_2$ layer is inserted between the bottom electrodes Cu and MgO, the leakage current of the device is significantly reduced. Therefore, Cu/MoS$_2$/MgO/Cu device has the highest commercial value from the point of view of practical applications. Finally, according to the XPS results and XRD results, we establish the conductive filament models for the three devices, and analyze the reasons for the different resistive switching characteristics of the three devices.

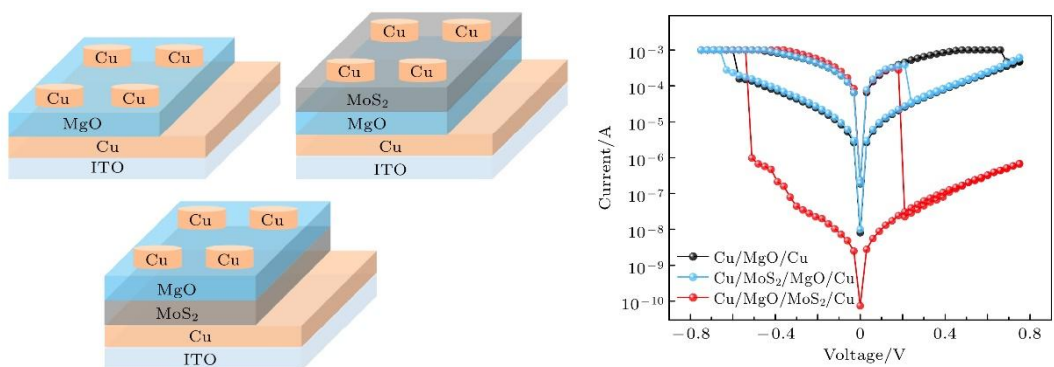

**Keywords:** resistive switching characteristic, interface barrier, switching ratio

PACS: 85.25.Hv, 81.07.–b, 73.40.Qv, 85.35.–p

## 1. Introduction

In the past few decades, with the rapid development of the Internet era, the amount of information is also rising, and there is an urgent need for new storage devices to store data [1–3]. Due to "Feng" · Prediction of "Neumann bottleneck" [4], researchers have turned their attention from traditional memory to the emerging random access memory (RAM), among which resistive random access memory (RRAM) has attracted much attention in recent years because of its high speed, low cost, high memory density and excellent scalability [5–7]. However, there are still problems such as device variability, power consumption in embedded applications, and read and write interference in crossbar array integration, which hinder the commercialization of RRAM [8–10]. So far, resistive switching (RS) has been detected in materials such as binary metal oxides [11], multicomponent oxides (especially those with perovskite structures) [12-14], and organic compounds [15]. In binary metal oxides, $HfO_2$ and $Ta_2O_5$ has been widely studied because of its good resistance switching performance and mature process technology [16–19]. For example, Kumar et al. [17] reported an ITO/MgO/$HfO_2$/ITO transparent resistive switching devices that exhibit photo response through defect engineering in the switching layer, which leads to subsurface active switching positions in the formed conductive filaments, thereby reducing oxygen loss through the polycrystalline electrode. The device has excellent switching ratios (~ $10^7$), highly stable DC setting and light setting durability (>1000 cycles without degradation), excellent retention (>$10^4$ s @ 85 °C), high transparency (>85% visible transmittance), and a light setting response time of 30 μs. Also, the enhancement of device performance due to the insertion of MgO layer was mentioned in this study. Lee et al. [19] proposed a bipolar resistance switching memory based on $TaO_x$, the structure of Ru/$Al_2O_3$/$Ta_2O_5$/$TaO_x$/$Al_2O_3$/W. The device has excellent memory performance, including fast operating speed (~ 10 ns), good switching endurance (~ $10^6$ Cycles) and stable data retention (>$10^4$ S @ 200 °C). In contrast, the research of MgO in RRAM needs more attention, and there are relatively few relevant references at present [20–22]. However, both MgO and $HfO_2$ belong to binary metal oxides and have many similarities in electrical and optical properties, such as high transmittance, wide band gap, and stable chemical properties [23]. At the same time, studies have shown that MgO has excellent resistance switching characteristics and is one of the potential candidate materials for RRAM application [24].

MgO is a common metal oxide with wide band gap (7.3-7.8 eV), moderate dielectric constant

(9.8-10), high thermal stability and high breakdown electric field (up to 12 MV/cm) [23]. Inorganic transition metal sulfide $MoS_2$ with variable band gap (bulk $MoS_2$ and monolayer $MoS_2$ are 1.27 eV and 1.98 eV, respectively), high electron affinity ($\chi \sim 4$ eV), large dielectric constant ($k \sim 4$-17) and chemical stability, it has attracted much attention as a switching layer material for RRAM in recent years [25–27]. However, there is a lack of literature on the resistive switching characteristics and current conduction mechanism of RRAM based on bilayer $MgO/MoS_2$.

In this paper, resistance switching characteristics of single-layer MgO device, double-layer $MoS_2/MgO$ stack device and bilayer $MgO/MoS_2$ stack device are studied. During the analysis, the low power consumption and large switching ratio of bilayer $MgO/MoS_2$ was observed. At the same time, the results of the *I-V* curves show that the main conduction mechanism of single-layer MgO device and the bilayer $MoS_2/MgO$ stack device is space-charge-limited conduction (SCLC), while in bilayer $MgO/MoS_2$ stack device, the regulation of the interface barrier is found, and the regulation of the interface barrier varies with the insertion position. In addition, the results of characterization test analysis support the proposed mechanism model and provide an explanation for the resistance switching characteristics of bilayer device.

## 2. Experiments and Methods

Cu/MgO/Cu, Cu/MgO/$MoS_2$/Cu and Cu/$MoS_2$/MgO/Cu multilayer films are deposited on three identical indium tin oxide (ITO) conductive glass substrates by magnetron sputtering at room temperature. The ITO substrate is commercially purchased. The size of a single ITO glass sheet is 20 mm×10 mm×1.1 mm, the thickness of the ITO film is about 185 nm, the square resistance of the ITO substrate is less than 6 Ω, and the transmittance is greater than 84%. Before sputtering, in order to ensure the quality of thin film growth, the background vacuum pressure of the cavity is $1.8 \times 10^{-4}$ Pa, and the flow rate of high purity argon is 30 SCCM(Standard cubic centimeter per minute) during sputtering, and keep the pressure of the cavity at 1 Pa during sputtering. The process of preparing Cu/MgO/$MoS_2$/Cu device is as follows: First, in the case of DC sputtering power of 30 W, 150 nm Cu is deposited on the surface of ITO as the bottom electrode (BE). Then, 120 nm $MoS_2$ was deposited on Cu with RF sputtering power of 50 W. Subsequently, 60 nm MgO was deposited on $MoS_2$ with a RF sputtering power of 30 W. Finally, in the case of DC sputtering power of 30 W, 150 nm Cu is deposited on the surface of MgO as the top electrode (TE). The other two devices were prepared in the same way, and Fig. 1(a) (b) (c) show the structural diagrams of the three devices, respectively. To analyze the electrical characteristics of the device, the *I-V* curve of the devices is measured using the Keithley 2400 source meter. At the same time, in order to determine the crystal phase of the thin film and the element types and chemical states contained in the device, the device was tested by X-ray diffractometer (XRD) and X-ray photoelectron spectroscopy (XPS). Among them, XPS and XRD test samples are MgO/MoS$_2$/Cu devices.

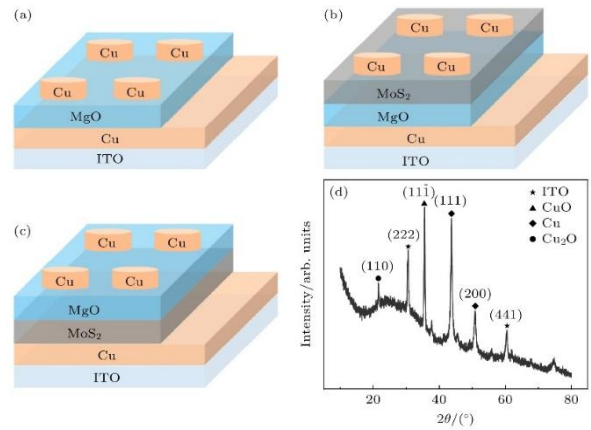

**Figure 1.** Structure diagram and XRD pattern of device: (a) Structure of Cu/MgO/Cu; (b) structure of Cu/MoS$_2$/MgO/Cu; (c) structure of Cu/MgO/MoS$_2$/Cu; (d) XRD pattern of MgO/MoS$_2$/Cu.

## 3. Results and Discussion

Fig. 1(d) is the XRD pattern of MgO/MoS$_2$/Cu device, where (2 2 2) and (4 4 1) are the crystallization peaks of ITO, (1 1 $\bar{1}$) is the crystallization peak of CuO, (1 1 1) and (2 0 0) are the crystallization peaks of Cu, and (1 1 0) is the crystallization peak of Cu$_2$O. At the same time, there is no obvious corresponding crystallization peak of MgO and MoS$_2$, which indicates that MgO and MoS$_2$ in the device should be amorphous [28]. It is well known that most of the films grown by magnetron sputtering at room temperature are amorphous [29]. At the same time, after referring to and comparing the XRD patterns from multiple related literatures, [12,13,16,17,30,31] found that there is no corresponding diffraction peak of MgO and Mos$_2$ in Fig. 1(d), which confirms that MgO and Mo$_2$ are amorphous. The XPS pattern of MgO/MoS$_2$/Cu device is shown in Fig. 2, where Fig. 2(a) is the auger electron spectroscopy (AES) of Cu after nonlinear least square fitting. The fitting results show that there is no metal Cu, and the Cu element mainly comes from CuO and Cu$_2$O. In order to further prove the chemical state of Cu in the sample, the Cu 2p spectrum was fitted, and the results are shown in Fig. 2(b). The Fig. 2(b) shows that Cu 2p can be divided into seven peaks, in which the peak at 962.79 eV corresponds to the strong satellite peak of Cu$^{2+}$, and there are two peaks at the positions of Cu 2p$_{1/2}$ and Cu 2p$_{3/2}$, corresponding to the peaks of CuO and Cu$_2$O, respectively. The peak positions of CuO are on the left, 954.47 eV and 934.63 eV, respectively, and the peak positions of Cu$_2$O are on the right, 952.53 eV and 932.74 eV, respectively. The remaining two peaks are 944 eV and 940.82 eV, which correspond to the mixed signals of Cu$^{2+}$ strong satellite peak and Cu$^+$ microsatellite peak. The XPS analysis of Cu shows that there is Cu oxide in the MgO film. Fig. 2(c) is the fitting result of Mg 1s spectrum, and it can be seen that the noise interference of the spectrum is large, which indicates that the signal of Mg element is weak, but there is still a weak peak at 1303.88 eV, corresponding to metal Mg, which may be related to the oxidation process of Cu. The Fig. 2(d) is the fitting result of the O 1s spectrum. The higher binding energy at about 531.8 eV corresponds to the oxygen vacancy ($V_O$) in the switching layer (SL), while the lower binding energy at about 530.64 eV corresponds to the lattice oxygen ion (Mg—O bond), which indicates that the migration of oxide ions may occur in the device and a new oxide is formed [23].

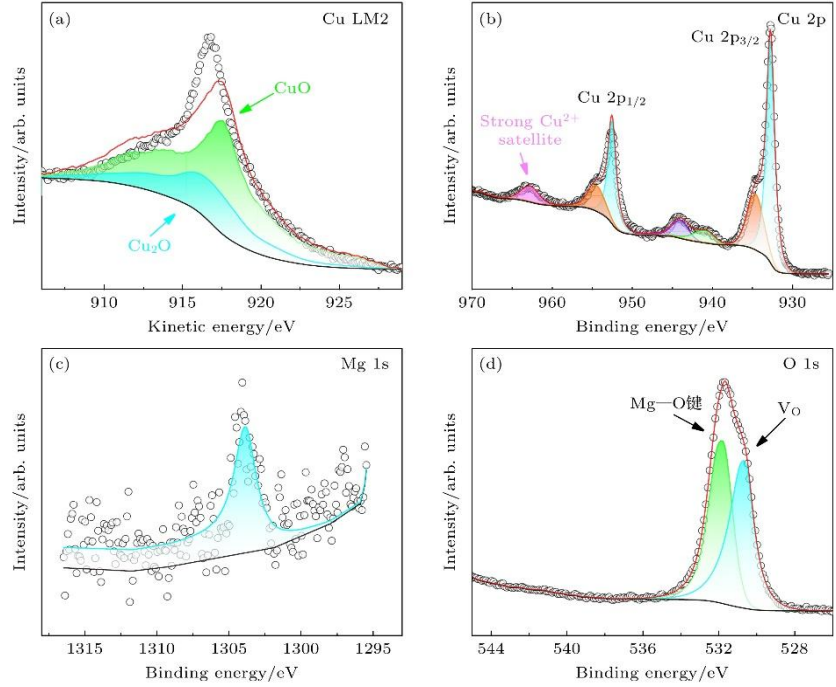

**Figure 2.** XPS spectra of the device: (a) AES of Cu; (b) XPS spectrum of Cu 2p; (c) XPS spectrum of Mg 1s; (d) XPS spectrum of O 1s

The Keithley 2400 source table was used to measure the *I-V* curve of the device, and the measurement path was 0 V→0.75 V→0 V→-0.75 V→0 V. The *I-V* curves of the three devices were plotted in semilog coordinates, and the results were shown in Fig. 3, where the right figure was the red cyclic curve highlighted in the left figure. Moreover, the cyclic curves of the three devices all conform to the *I-V* curve of the Counter-figure-of-eight (cF8) bipolar switching, and the original devices all need to undergo soft breakdown (Forming or Electroforming) process [32]. By comparing the *I-V* curves of Cu/MgO/Cu and Cu/MoS$_2$/MgO/Cu devices, it is observed that the reset voltage ($V_{Reset}$) of the Cu/MoS$_2$/MgO/Cu device is significantly lower than that of the Cu/MgO/Cu device, while their set voltages ($V_{Set}$) remain comparable. This indicates that the inserted MoS$_2$ layer reduces the power consumption of the Cu/MgO/Cu device. Similarly, by comparing the *I-V* curves of Cu/MgO/Cu and Cu/MgO/MoS$_2$/Cu, it can be found that the switching ratio of Cu/MgO/Cu devices is about 10, but after MoS$_2$ layer is inserted, the switching ratio of Cu/MgO/MoS$_2$/Cu devices is about $10^3$. In addition, the $V_{Reset}$ of Cu/MgO/MoS$_2$/Cu devices decreases greatly, and the $V_{Set}$ decreases slightly, which indicates that the switching ratio of the devices is improved to a certain extent after $MoS_2$ layer is inserted, and the overall power consumption of the devices is greatly reduced. Finally, by comparing the *I-V* curves of $Cu/MoS_2/MgO/Cu$ and $Cu/MgO/MoS_2/Cu$, it can be found that the insertion position of $MoS_2$ layer mainly affects the leakage current, $V_{Reset}$ and $V_{Set}$ of the device. This shows that the $MoS_2$ layer inserted between the bottom electrode and MgO can effectively improve the switching ratio of the device and reduce the power consumption of the device, but the $MoS_2$ layer inserted between the top electrode and MgO is basically ineffective to improve the switching ratio of the device. In order to compare the performance of the three devices more intuitively, the comparison diagram shown in Fig 4 is drawn. It can be clearly found that $Cu/MgO/MoS_2/Cu$ has the largest switching ratio, and both $V_{Reset}$ and $V_{Set}$ are the smallest. The main difference between $Cu/MgO/Cu$ and $Cu/MoS_2/MgO/Cu$ is the $V_{Reset}$ voltage.

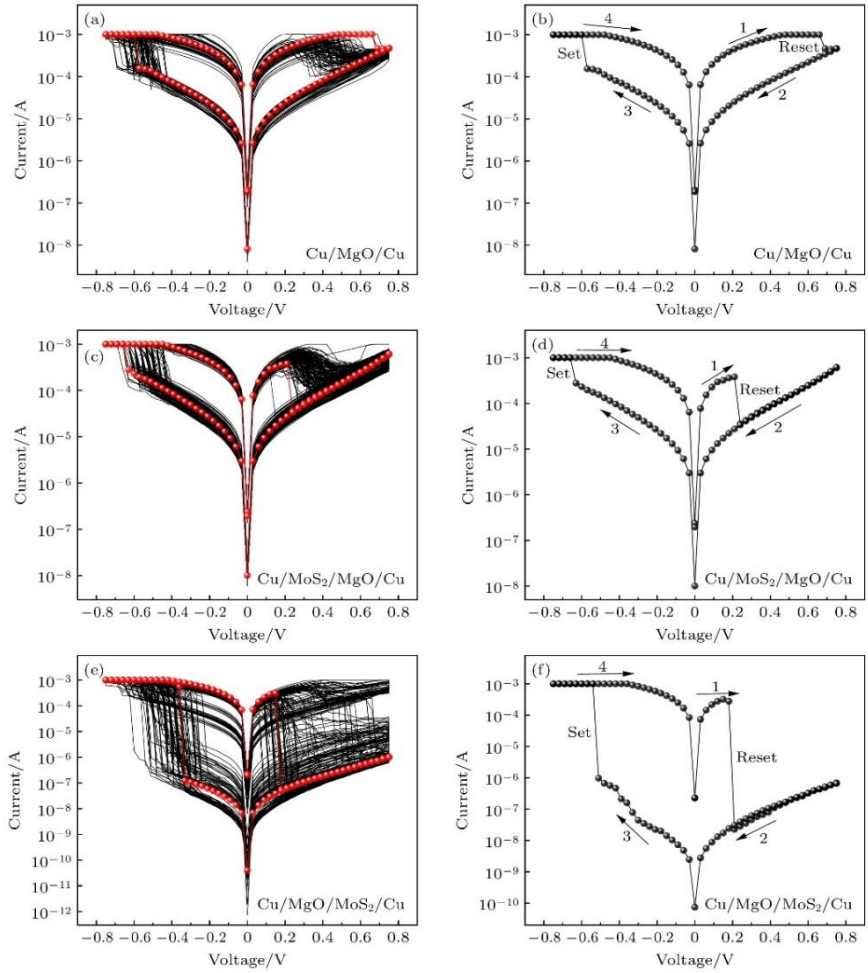

**Figure 3.** *I-V* cycle trajectories of the three devices in semilog coordinates: (a) Multiple cycle curves of $Cu/MgO/Cu$; (b) single cycle curve of $Cu/MgO/Cu$; (c) multiple cycle curves of $Cu/MoS_2/MgO/Cu$; (d) single cycle curve of $Cu/MoS_2/MgO/Cu$; (e) multiple cycle curves of $Cu/MgO/MoS_2/Cu$; (f) single cycle curve of $Cu/MgO/MoS_2/Cu$.

In order to further compare the reliability of the three devices, the cumulative probability distribution maps of $V_{Reset}$ and $V_{Set}$ of the three devices are drawn respectively, so as to compare the dispersion of $V_{Reset}$ and $V_{Set}$, and the results are shown in Fig. 5. By observing the Fig. 5, it can be found that the $V_{Reset}$ and $V_{Set}$ of the three devices are relatively concentrated. Relatively speaking, the $V_{Reset}$ and $V_{Set}$ of the Cu/MgO/MoS$_2$/Cu device have the highest reliability, especially the $V_{Reset}$. In addition, the distribution of high and low resistance states of the three devices is shown in Fig. 6. The Cu/MgO/Cu device underwent approximately 120 consecutive set/reset cycles, the Cu/MoS$_2$/MgO/Cu device underwent approximately 180 consecutive set/reset cycles, and the Cu/MgO/MoS$_2$/Cu device underwent approximately 200 consecutive set/reset cycles. Combined with the cycling trajectory of Fig. 3, it can be seen that the Cu//MoS$_2$/MgO/Cu device has the best cycling tolerance and stability among the three devices, but the switching ratio is small (about 10). At the same time, the stability of the high resistance state (HRS) of the Cu/MgO/MoS$_2$/Cu device is relatively poor, and the resistance distribution before and after 40 cycles is quite different, while the low resistance state (LRS) is relatively stable, and the switching ratio of the device fluctuates between 10 and $10^4$, but in general, the switching ratio is concentrated around $10^3$, which indicates that the MoS$_2$ insertion layer will affect the resistance distribution of the device HRS and improve the switching ratio of the device, but the HRS stability of the device is reduced.

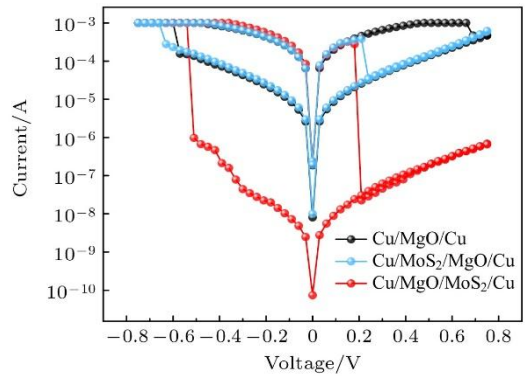

**Figure 4.** Single cycle comparison diagram of the three devices in semilog coordinates.

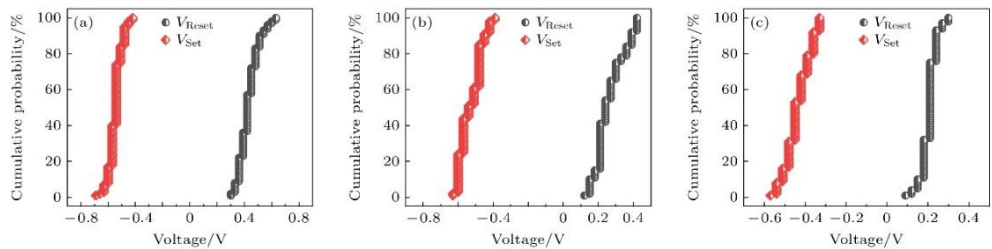

**Figure 5.** Cumulative probability distribution of $V_{Reset}$ and $V_{Set}$ for the three devices: (a) Cu/MgO/Cu; (b) Cu/MoS$_2$/MgO/Cu; (c) Cu/MgO/MoS$_2$/Cu.

In order to explore the above experimental phenomena and the switching mechanism of the devices, the *I-V* curves of the three devices are linearly fitted in log-log coordinates to determine the differences of the switching mechanism of the three devices. The Fig. 7(a) (c) (e) are the fitting results under positive bias voltage, and the Fig. 7(b) (d) (f) are the fitting results under negative bias voltage. The red part in the figure is LRS, and the blue and green parts are HRS. By observing the Fig. 7, it can be seen that the fitting curve of LRS presents a straight line with a slope of 1, indicating that the conduction mechanism is mainly ohmic conduction, and the fitting curve of HRS is similar at low voltage, which also corresponds to the ohmic conduction model. As the voltage increases, the curve exhibits a linear relationship with a slope close to 2, which indicates that the device follows the classical SCLC. The classical SCLC consists of three regions: the low voltage ohmic region (*I-V*), the Child square region (*I-V$^2$*), and the current ramp region (*I-V$^{>2}$*) [28]. In summary, the conduction mechanisms of the three devices are consistent with SCLC as a whole, but the fitting results of the high voltage region of the bilayer device are not fully consistent with SCLC, which can be attributed to the regulation of the MgO/MoS$_2$ interface barrier.

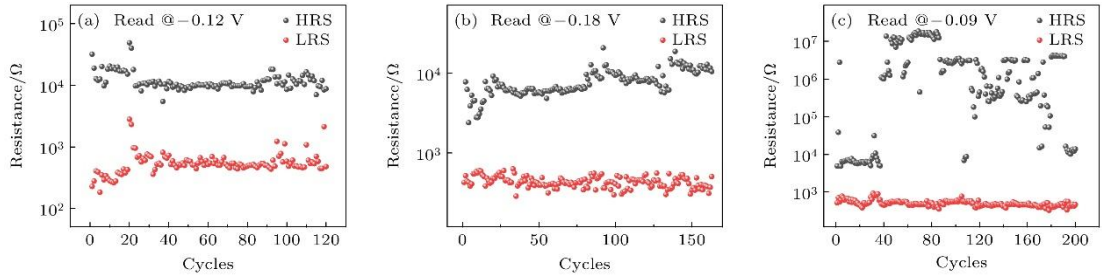

**Figure 6.** HRS and LRS distribution of the three devices: (a) Cu/MgO/Cu; (b) Cu/MoS$_2$/MgO/Cu; (c) Cu/MgO/MoS$_2$/Cu.

All three devices undergo a soft breakdown process (which occurs at a negative voltage) before the normal cycling process. In the XRD results, the crystallization peak of CuO was found, and the peak of Cu oxide (CuO$_x$) was also found in the XPS results, but the test sample was MgO/MoS$_2$/Cu, and the Cu element should come from the bottom electrode, which indicates that a small amount of Cu has been dissolved into the MgO film on the surface during the magnetron sputtering process, but the content is low, which is consistent with the data results. Therefore, the resistance switching mechanism of the three devices can be attributed to the formation and breakage [33] of the Cu conductive filament (CF), and the principle schematic diagrams of the reset process and the set process of the devices are shown in Fig. 8. The oxidation and reduction reactions of Cu are as follows:

$$Cu \rightarrow Cu^{n+} + ne^- \tag{1}$$

$$Cu^{n+} + ne^- \rightarrow Cu \tag{2}$$

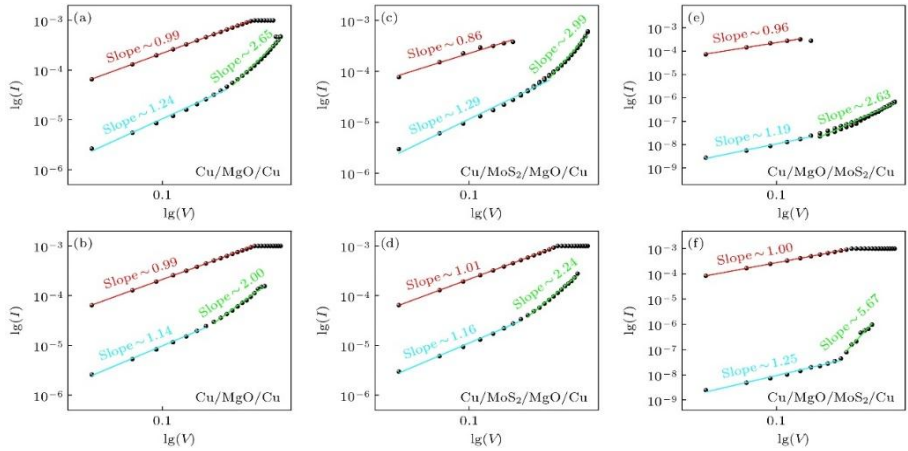

**Figure 7.** *I-V* curves fitting diagram of the three devices in double logarithm coordinates: (a) Fitted curve of Cu/MgO/Cu at positive voltage; (b) fitted curve of Cu/MgO/Cu at negative voltage; (c) fitted curve of Cu/MoS$_2$/MgO/Cu at positive voltage; (d) fitted curve of Cu/MoS$_2$/MgO/Cu at negative voltage; (e) fitted curve of Cu/MgO/MoS$_2$/Cu at positive voltage; (f) fitted curve of Cu/MgO/MoS$_2$/Cu at negative voltage.

The peaks of Mg 1s and V$_O$ measured in XPS are both corresponding to Cu oxides, but they are also trace. Before the reset process, the device had a soft breakdown process under negative voltage, and the BE Cu was oxidized to form e$^-$ and Cu$^{n+}$, which migrated to the positive and negative electrodes respectively under the action of the electric field. When the Cu$^{n+}$ migrated to the negative electrode, the e$^-$ near the negative electrode was reduced to Cu, but a small amount of Cu$^{n+}$ in the MgO layer had formed CuO$_x$ with a small amount of O$^{2-}$ migrating to the positive electrode, which did not reach the negative electrode, which was consistent with the XPS results. Because MgO has a wide band gap and can be considered as an insulator, it is very difficult for carriers to pass through the MgO layer, and a high Schottky barrier will be formed between MgO and TE Cu. These factors lead to the possibility that the narrow part of the Cu conductive filament is at the interface between TE Cu and MgO [34–36]. Of course, if it is the Cu/MoS$_2$/MgO/Cu structure in Fig. 8(c) and Fig. 8(d), ions need to migrate through a layer of MoS$_2$ medium after passing through the MgO layer, so the narrow part of the filament is likely to be at the interface between TE Cu and MoS$_2$ Meanwhile, based on the above reasons, the shape of the filament is likely to be the inverted cone shape shown in Fig. 8 [23].

During the reset process, the $V_{Reset}$ of the single-layer device is obviously different from that of the double-layer device. As shown in the schematic diagram of Fig. 8(a) and Fig. 8(b), under positive voltage, Cu near the positive electrode is oxidized to form Cu$^{n+}$, which migrates to the negative electrode under the action of the electric field, resulting in the fracture of CF at the interface between TE Cu and MgO, which is the red shaded part in the figure. However, due to the barrier of only the MgO layer in the middle, the CF width of the single-layer device may be larger, and a larger voltage is required to break the CF from the narrow part, so that the device returns from the LRS to the HRS. On the contrary, in the bilayer device, the CF is more difficult to form under the action of two barrier layers, and the width of the filament is smaller, so the voltage required for fracture is smaller.

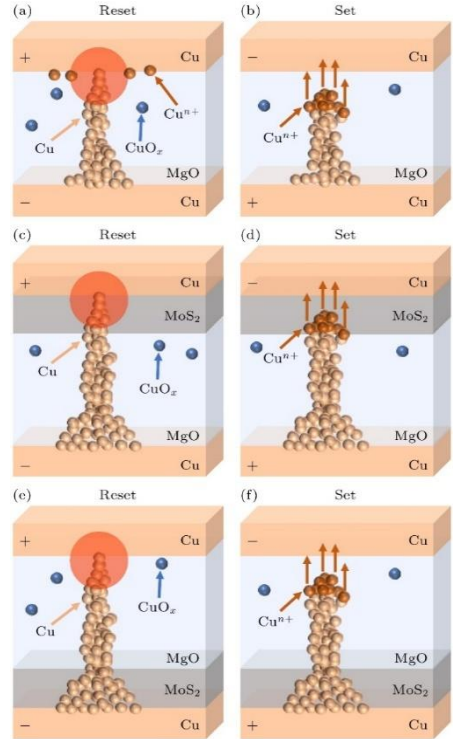

**Figure 8.** Schematic diagram of the mechanism explanation of the three devices: (a) Reset process of Cu/MgO/Cu; (b) set process of Cu/MgO/Cu; (c) reset process of Cu/MoS$_2$/MgO/Cu; (d) set process of Cu/MoS$_2$/MgO/Cu; (e) reset process of Cu/MgO/MoS$_2$/Cu; (f) set process of Cu/MgO/MoS$_2$/Cu.

In the setting process, the mechanism of the three devices is basically the same. First, under negative voltage, the Cu$^{n+}$ formed by Cu oxidation begins to migrate to the negative electrode and is reduced at the narrow CF, reconnecting the broken CF and switching the device from HRS to LRS. At the same time, it should be noted that the current of Cu/MgO/MoS$_2$/Cu increases very fast near $V_{Set}$ and, which may be related to the narrow position of the filament and the regulation of the interface barrier.

On the other hand, it is noted that while all three devices exhibit symmetric structures, only the single-layer device demonstrates symmetric *I−V* curves [37]. Based on this, the differences between Cu/MoS$_2$/MgO/Cu and Cu/MgO/MoS$_2$/Cu are emphasized. The main difference between the two is the position of the MoS$_2$ layer, which leads to the difference in leakage current, that is the resistance value of the HRS, which can be attributed to the regulatory role of the MgO/MoS$_2$ interface barrier [38]. Carrier migration from MgO to MoS$_2$ is easier than the reverse process, mainly due to the larger bandgap of MgO compared to MoS$_2$, as well as the generally higher work function of amorphous MgO. This results in a higher potential barrier on the MgO side relative to the MoS$_2$ side, leading to differences in the required energy (manifested as voltage differences in the device). This asymmetry in potential barriers is the primary cause of the asymmetric *I−V* curves in bilayer devices [39]. Additionally, the HRS of the two devices differs after the reset process. As shown in Fig. 8(c) and 8(e), the reset process for Cu/MoS$_2$/MgO/Cu Cu/MoS$_2$/MgO/Cu involves the migration of Cu$^{n+}$ from MoS$_2$ to MgO, which is more challenging. In contrast, the reset process for Cu/MgO/MoS$_2$/Cu involves Cu$^{n+}$ migration from MgO to MoS$_2$, which is relatively easier. This results in a significantly reduced leakage current and increased HRS resistance in the Cu/MgO/MoS$_2$/Cu device. However, the reduced migration difficulty also leads to a substantial increase in HRS randomness, ultimately causing the scattered HRS results observed in Fig. 6(c).

**4. Conclusion**

In conclusion, three types of stacked structures based on MgO were prepared by magnetron sputtering, and it was found that the MoS$_2$ insertion layer has a great influence on the resistance switching performance of single-layer MgO devices, which is related to the insertion position. The Cu/MgO/MoS$_2$/Cu device has the best switching ratio. By analyzing the results of XPS characterization and the *I-V* curve of the device, the effect of MoS$_2$ insertion layer on the resistance switching characteristics of Cu/MgO/Cu device is mainly attributed to the control of Cu conductive filaments and interface barriers. On the other hand, in terms of reliability and stability, the Cu/MoS$_2$/MgO/Cu device performs the best, meeting the commercial value and basic standards of resistive random access memory. In our previous work, we found that the thickness change of MoS$_2$ has a great influence on the resistance switching characteristics of BTO/MoS$_2$ stack devices, and the increase of thickness will lead to more defect States, thus reducing the resistance [40] of the high resistance state of the device. At the same time, we investigated the relevant literature, and found that the influence of the thickness of MgO layer on the resistance switching characteristics of the device is usually related to the formation voltage, and has little influence on the overall performance of the device, and there are few reports in the [38,41]. In the later work of this study, we will investigate the influence rules of the ratio or thickness changes of MgO and MoS$_2$ in the Cu/MgO/MoS$_2$/Cu device on the device performance.

In this paper, the effect of MoS$_2$ insertion layer on the resistance switching performance of Cu/MgO/Cu device is investigated. On the other hand, the traditional mechanism explanation of MgO-based resistance switching devices generally only introduces the conductive filament model, while the research work in this paper shows that the interface barrier also has a regulatory effect on the resistance switching characteristics of the device. This paper has a certain reference value for the research of MgO in the field of RRAM, and provides a new idea for the research of MgO/MoS$_2$ stack device.

## Data availability statement

The linked data of this paper can be accessed in thehttps://doi.org/10.57760/sciencedb.j00213.00086 of the Scientific Data Bank.